\documentclass[12pt]{article}
\usepackage{amssymb, latexsym, amsmath}
\usepackage[plainpages=false,pdfpagelabels,pdftex,bookmarksnumbered,colorlinks,backref, bookmarks, breaklinks, linktocpage]{hyperref}
\usepackage[pdftex]{graphicx}

\begin{document}

\title{\textbf{Pseudoduality and Complex Geometry in Sigma Models}}

\author{\textbf{Mustafa Sarisaman} \footnote{E-mail address: msarisaman@ku.edu.tr, Phone: +90 212 338 1378, Fax: +90 212 338 1559.}}

\date{}

\maketitle
\par
\begin{center}
\textit{Department of Mathematics\\ Ko\c{c} University \\ 34450 Sar{\i}yer\\ Istanbul, Turkey }
\end{center}

\

\begin{center}
Monday, October 1, 2012
\end{center}

\

\begin{abstract}
We study the pseudoduality transformations in  two dimensional $N = (2, 2)$ sigma models on K\"{a}hler manifolds. We show that structures on the target space can be transformed into the pseudodual manifolds by means of (anti)holomorphic preserving mapping. This map requires that torsions related to individual spaces and riemann connection on pseudodual manifold  must vanish. We also consider holomorphic isometries which puts additional constraints on the pseudoduality.
\end{abstract}

\vfill

\section{Introduction}\label{sec:int}

In this paper we are going to analyze the pseudoduality
transformations \cite{alvarez1, alvarez2, curtright1, ivanov} in
sigma models in the realm of complex geometry. We will focus on $N =
2$ supersymmetric sigma models to achieve this goal, and reveal the
constraints arising from complex structures. The main motivation of
this study is to find out a way to perform pseudoduality
transformations in topological sigma models which are obtained by
twisting $N = 2$ supersymmetric sigma models, and to see their
effects on topological invariants of the pseudodual models.

Non-linear sigma models in two dimensions form majority of current
studies in string theory and condensed matter physics by virtue of
their rich mathematical structures. It appears that their
world-sheet supersymmetric extensions attract much of interest
especially in understanding their target space geometries. It has
been revealed \cite{ketov1} that target space of $N = 1$ sigma
models is a (pseudo-)Riemannian manifold, $N = 2$ is the K\"{a}hler
manifold and $N = 4$ is the hyper-K\"{a}hler manifold. That is why
we consider our sigma models based on K\"{a}hler manifolds.

It is well-known that usual duality transformation is performed on
the target space and preserves the Hamiltonian. However,
pseudoduality transformation is established on the
worldsheet\footnote{This is the origin of the nomenclature
``pseudo''.} and preserves the stress-energy tensor. Therefore, this
transformation is not a canonical transformation. Rather, this is
``on shell'' duality transformation which maps solutions of the
equations of motion of the ``pseudodual'' models. We refer the term
``pseudodual'' if there is a pseudoduality transformation between
different models. In paper \cite{alvarez1} pseudodulity in classical
sigma models was studied, and we generalized this constructions to
$N = 1$ supersymmetric sigma models \cite{msarisaman1}. We figured
out that pseudoduality transformation restricts the target space
geometry to vanishing torsion in $N = 1$ case. Moreover, the
manifolds on which sigma models are based were found to be dual
symmetric spaces with opposite curvatures relative to each other.

It has been pointed out \cite{zumino} that $N = 2$ supersymmetric
sigma model with chiral superfields in two dimensions is represented
by a target space which has K\"{a}hler geometry. As known by `Gates,
Hull and Rocek theorem', this geometry is bi-hermitian which has two
complex structures, and the metric is hermitian with respect to both
of them. In what follows we first outline the basics of K\"{a}hler
geometry and then progress to construct the pseudoduality
transformations on this geometry to find out the necessary
conditions.

\section{K\"{a}hler Geometry and Sigma Models}\label{sec:KGSM}

In this section we review the K\"{a}hler geometry which arises in the study of sigma models, especially $N = (2,2)$ supersymmetric case for our purpose. The scalar fields in 2 dimensional $N = 2$ supersymmetry can be described by the following off-shell constraint relations

\begin{equation}
\mathbb{\bar{D}}_{\pm} \Phi = 0 , \ \ \ \ \ \ \mathbb{D}_{\pm} \bar{\Phi} = 0 \label{equation1}
\end{equation}
where $\mathbb{D}_{\pm}$ and $\mathbb{\bar{D}}_{\pm}$ are the $N = (2, 2)$ chiral and antichiral superspace covariant derivatives respectively, and $\Phi$ and $\bar{\Phi}$ are the chiral and antichiral superfields. The ordinary field components of the chiral superfield $\Phi$ in terms of $N = (1, 1)$ fields are

\begin{equation}
x = \Phi | \ \ \ \ \ \psi_{\pm} = \mathbb{D}_{\pm} \Phi | \ \ \ \ \ F = \mathbb{D}_{-} \mathbb{D}_{+}\Phi | \notag
\end{equation}
where $|$ represents $\theta$-independent component of a superfield, $x$ is the scalar bosonic field, $\psi_{\pm}$ is the two dimensional Majorana spinor field, and $F$ is the auxiliary scalar field. They form the $N = (1, 1)$ superfield $X$ as
\begin{equation}
X = x + \theta^{+} \psi_{+} + \theta^{-} \psi_{-} + \theta^{+} \theta^{-} F \notag
\end{equation}

The most general $N = (2, 2)$ supersymmetric action in terms of a single function $K(\Phi, \bar{\Phi})$ called the K\"{a}hler potential is written as

\begin{equation}
S = \int d^{2} \sigma d^{2} \theta d^{2} \bar{\theta} K(\Phi, \bar{\Phi}) \label{equation2}
\end{equation}

The constraint relations (\ref{equation1}) reduce this action to
\begin{equation}
S = - \frac{i}{2} \int d^{2} \sigma d^{2} \theta \ g_{AB} D_{+} X^{A} D_{-} X^{B} \label{equation3}
\end{equation}
where $D_{\pm}$ is the supercovariant derivative
\footnote{Supercovariant derivative of $N=1$ supersymmetry is given
by $D_{\pm} =
\partial_{\theta^{\pm}} + i \theta^{\pm}
\partial_{\pm}$, see \cite{ketov1, msarisaman1}.} corresponding to $N = 1$ supersymmetry and the
metric $g_{AB}$ is written as
\begin{equation}
g_{AB} = \begin{pmatrix}
           0 & K_{a\bar{b}} \\
           K_{\bar{a}b} & 0 \\
         \end{pmatrix} \label{equation4}
\end{equation}
We will use the indices $a, b, c ...$ to denote the holomorphic
coordinates, and $\bar{a}, \bar{b}, \bar{c}, ...$ the
anti-holomorphic ones. The vector $X^{A}$ can be split into $X^{a}$
and $\bar{X}^{\bar{a}}$, $\{A\}$ represents the collection of
holomorphic and anti-holomorphic indices, and $K_{a\bar{b}}$ is
defined by $K_{a\bar{b}} := \frac{\partial}{\partial X^{a}}
\frac{\partial}{\partial X^{\bar{b}}} K$. As will be seen below the
target space geometry is K\"{a}hler with metric $g_{\mu\bar{\nu}} =
\partial_{\mu} \partial_{\bar{\nu}} K$.

\par We would like to extract the K\"{a}hler structures using $N = (1, 1)$ supersymmetry. It is noted that second supersymmetry acts on the fields as
\begin{equation}
\delta^{2} (\epsilon) X^{i} = \epsilon^{+} J_{(+)k}^{i} D_{+} X^{k} + \epsilon^{-} J_{(-)k}^{i} D_{-} X^{k} \notag
\end{equation}
where $J_{(\pm)k}^{i}$ are defined as the almost complex structures in the target space. The action is invariant under this second supersymmetry if
\begin{align}
J_{(\pm)ij} &= - J_{(\pm)ji} \label{equation5}\\
\nabla^{\pm} J_{(\pm)} &= 0 \label{equation6}
\end{align}
where the covariant derivative $\nabla^{\pm}$ has connection $\Gamma_{jk}^{\pm i} = \Gamma_{jk}^{i} \pm g^{il}H_{ljk}$. These tensors obey the following conditions
\begin{align}
J_{(\pm)k}^{i} J_{(\pm)j}^{k} &= - \delta_{j}^{i} \notag\\
N_{jk}^{i} = J_{(\pm)l}^{i} \partial_{[k}J_{(\pm)j]}^{l} &+ \partial_{l} J_{(\pm)[k}^{i} J_{(\pm)j]}^{l} = 0 \notag
\end{align}
where $N_{jk}^{i}$ is called the Nijenhuis tensor, and the fact that it vanishes implies that complex structures are integrable. It is said that target space of real dimension $2n$ covered by a system of coordinate neighborhoods $X^{i}$ is equipped with an almost complex structures $J_{(\pm)}$. We drop the $(\pm)$ indices, and define the projectors
\begin{equation}
P_{\pm} = \frac{1}{2} (1 \pm i J) \label{equation7}
\end{equation}
to split any vector $V^{i}$ into projections $V_{\pm}^{i}$. If we describe the basis 1-forms by $dX^{i}$ we may define the fundamental two-form
\begin{equation}
w = J_{ij} dX^{i} \wedge dX^{j} \notag
\end{equation}
which shows that target space manifold is Hermitian manifold. A K\"{a}hler manifold is obtained by a Hermitian manifold with a closed fundamental two-form, i.e., $dw = 0$, or equivalently $J_{[ij, k]} = 0$. It is obvious that $N = (2, 2)$ supersymmetric sigma model in two dimensions is endowed with the target space a K\"{a}hler geometry once reduced to $N = (1, 1)$ supersymmetric sigma model with target space a riemannian manifold of real dimension 2n by imposing K\"{a}hler structure J.

\section{Pseudoduality on K\"{a}hler Manifolds}\label{sec:PKM}

Sigma model is identified by the map $\Phi: \Sigma \rightarrow
\mathcal{M}^{\mathbb{C}}$, where $\Sigma$ is the complexified
superspace (i.e. worldsheet extended by N=2 supersymmetry). We start
with the reduced action (\ref{equation3}), and let our manifold
$\mathcal{M}^{\mathbb{C}}$ \footnote{$\mathcal{M}^{\mathbb{C}}$ is
the complexified manifold $\mathcal{M}$, and can be decomposed into
holomorphic $\mathcal{M}^{+}$ and anti-holomorphic $\mathcal{M}^{-}$
parts by applying projection operators (\ref{equation7}).} involve
an antisymmetric two-form field $b_{AB}$. It is manifest that the
form of the metric only allows the connections $\Gamma_{bc}^{a} \in
\mathcal{M}^{+}$ and $\Gamma_{\bar{b}\bar{c}}^{\bar{a}} \in
\mathcal{M}^{-}$. Accordingly these connections are accompanied by
the corresponding torsions $H_{bc}^{a} \in \mathcal{M}^{+}$ and
$H_{\bar{b}\bar{c}}^{\bar{a}} \in \mathcal{M}^{-}$ where $H_{abc} =
\frac{1}{2}(\partial_{a}b_{bc} + \partial_{b}b_{ca} +
\partial_{c}b_{ab})$. It is noted that two-form field $b_{AB}$ on
manifold $\mathcal{M}^{\mathbb{C}}$ is split into components where
$b_{ab}$ and $b_{\bar{a}\bar{b}}$ vanish. This can be demonstrated
\cite{nakahara} with the condition that
\begin{equation}
b_{p} (X, Y) = b_{p} (J_{p}X, J_{p}Y) \notag
\end{equation}
at each point $p$ $\in$ $\mathcal{M}^{\mathbb{C}}$ and for any $X, Y
\in T_{p} \mathcal{M}^{\mathbb{C}}$. Therefore, the components of
the two-form field $b_{AB}$ will locally be expressed by
\begin{equation}
b_{AB} = \left(
           \begin{array}{cc}
             0 & b_{a\bar{b}} \\
             b_{\bar{a}b} & 0 \\
           \end{array}
         \right) \notag
\end{equation}
We next find out the equations of motion as follows
\begin{align}
X_{+-}^{c} = (\Gamma_{ab}^{c} - H_{bc}^{a}) X_{+}^{a} X_{-}^{b} \label{equation8}\\
\bar{X}_{+-}^{\bar{c}} = (\Gamma_{\bar{a}\bar{b}}^{\bar{c}} - H_{\bar{b}\bar{c}}^{\bar{a}}) \bar{X}_{+}^{\bar{a}} \bar{X}_{-}^{\bar{b}} \label{equation9}
\end{align}
where we defined $X_{\pm} := D_{\pm} X$, $X_{+-} := D_{-} D_{+} X$
and $\Gamma_{bc}^{a} := (K^{-1})_{d}^{a} K_{bc}^{d}$. We realize
that these equations are the equations of motion defined on the
holomorphic $\mathcal{M}^{+}$ and anti-holomorphic $\mathcal{M}^{-}$
parts of the target space when acted by the projectors
(\ref{equation7}) on $\mathcal{M}^{\mathbb{C}}$.

We would like to inquire about the pseudoduality conditions and hence write the corresponding pseudoduality equations. It is best to perform the analysis on the orthonormal coframes defined on $SO(\mathcal{M}^{\mathbb{C}})$ \footnote{$SO(\mathcal{M}^{\mathbb{C}})$ = $\mathcal{M}^{\mathbb{C}}$ $\times$ $SO(n)$. }. We choose an orthonormal frame $\{\Lambda^{a}\}$ with the connection one-form $\{\Lambda_{b}^{a}\}$ on the holomorphic part of superspace. We define the superspace by $z = (\sigma^{\pm}, \theta^{\pm})$. We may write the associated expressions for the anti-holomorphic parts by using bar on each item. The one-forms are given by
\begin{equation}
\Lambda^{a} = dz^{M} X_{M}^{a} \ \ \ \ \ \ \ \ \ \ \bar{\Lambda}^{\bar{a}} = d\bar{z}^{M} \bar{X}_{M}^{\bar{a}} \label{equation10}
\end{equation}
The covariant derivatives of $X_{M}$ and $\bar{X}_{M}$ defined in
$SO(\mathcal{M}^{\mathbb{C}})$ are
\begin{equation}
dX_{M}^{a} + \Lambda_{b}^{a} X_{M}^{b} = dz^{N} X_{MN}^{a} \ \ \ \ \ \ \ \ \ \ d\bar{X}_{M}^{\bar{a}} + \Lambda_{\bar{b}}^{\bar{a}} \bar{X}_{M}^{\bar{b}} = d\bar{z}^{N} \bar{X}_{MN}^{\bar{a}} \label{equation11}
\end{equation}
The Cartan structural equations will be
\begin{align}
d\Lambda^{a} = - \Lambda_{b}^{a} \wedge \Lambda^{b} \ \ \ \ \ d\Lambda_{b}^{a} = - \Lambda_{c}^{a} \wedge \Lambda_{b}^{c} + \Omega_{b}^{a} \label{equation12}\\
d\bar{\Lambda}^{\bar{a}} = - \bar{\Lambda}_{\bar{b}}^{\bar{a}} \wedge \bar{\Lambda}^{\bar{b}} \ \ \ \ \ d\bar{\Lambda}_{\bar{b}}^{\bar{a}} = - \bar{\Lambda}_{\bar{c}}^{\bar{a}} \wedge \bar{\Lambda}_{\bar{b}}^{\bar{c}} + \bar{\Omega}_{\bar{b}}^{\bar{a}} \label{equation13}
\end{align}
where $\Omega_{b}^{a}$ and $\bar{\Omega}_{\bar{b}}^{\bar{a}}$ are the curvature two-forms, and defined by $\Omega_{b}^{a} = \frac{1}{2}R_{b\bar{c}d}^{a} \bar{\Lambda}^{\bar{c}} \wedge \Lambda^{d} + \frac{1}{2}R_{bc\bar{d}}^{a} \Lambda^{c} \wedge \bar{\Lambda}^{\bar{d}}$ and $\Omega_{\bar{b}}^{\bar{a}} = \frac{1}{2}R_{\bar{b}\bar{c}d}^{\bar{a}} \bar{\Lambda}^{\bar{c}} \wedge \Lambda^{d} + \frac{1}{2}R_{\bar{b}c\bar{d}}^{\bar{a}} \Lambda^{c} \wedge \bar{\Lambda}^{\bar{d}}$.  It is apparent that pseudoduality equations may appear in forms that may either respect holomorphism of target spaces, or mix them by forming the mixture of structures. We first investigate the holomorphic pseudodaulity relations.

\subsection{Holomorphic Pseudoduality}\label{sec:HP}

This is the case where the (anti)holomorphic vectors are mapped to the associated (anti)holomorphic counterparts on the pseudodual manifold. Pseudoduality equations are
\begin{align}
\tilde{X}_{\pm}^{a} = \pm \mathcal{T}_{b}^{a} X_{\pm}^{b} \label{equation14}\\
\tilde{\bar{X}}_{\pm}^{\bar{a}} = \pm \mathcal{\bar{T}}_{\bar{b}}^{\bar{a}} \bar{X}_{\pm}^{\bar{b}} \label{equation15}
\end{align}
where $\mathcal{T} : T\mathcal{M}^{+} \longrightarrow
T\mathcal{\tilde{M}}^{+}$ and $\mathcal{\bar{T}} : T\mathcal{M}^{-}
\longrightarrow T\mathcal{\tilde{M}}^{-}$ are mappings between
associated tangent bundles with compatible holomorphicity. We
examine the integrability conditions which lead to above mentioned
pseudoduality transformations. Following the method employed in
\cite{msarisaman1} we take the exterior derivatives of the above
equations and use (\ref{equation11}) together with the convenient
use of (\ref{equation10}), (\ref{equation12}) and (\ref{equation13})
to obtain the results $H = \tilde{H} = \tilde{\Gamma} = 0$ (with all
holomorphic and anti-holomorphic indices) together with the
constraint relations
\begin{align}
d\mathcal{T}_{b}^{a} - 2 \mathcal{T}_{c}^{a} \Lambda_{b}^{c} + \tilde{\Lambda}_{c}^{a} \mathcal{T}_{b}^{c} = 0 \label{equation16}\\
d\mathcal{\bar{T}}_{\bar{b}}^{\bar{a}} - 2 \mathcal{\bar{T}}_{\bar{c}}^{\bar{a}} \bar{\Lambda}_{\bar{b}}^{\bar{c}} + \tilde{\bar{\Lambda}}_{\bar{c}}^{\bar{a}} \mathcal{\bar{T}}_{\bar{b}}^{\bar{c}} = 0 \label{equation17}
\end{align}
where the connection one-forms are defined as $\Lambda_{b}^{a} =
\Gamma_{bc}^{a} \Lambda^{c}$ and $\bar{\Lambda}_{\bar{b}}^{\bar{a}}
= \Gamma_{\bar{b}\bar{c}}^{\bar{a}} \bar{\Lambda}^{\bar{c}}$. Notice
that these equations are characteristic feature of pseudoduality,
not only special to sigma models. These equations contain
information about the geometry of the spaces on which pseudoduality
transformation is built. To reveal this information further
integrability conditions are required. Therefore the integrability
of (\ref{equation16}) and (\ref{equation17}) leads to the associated
curvature two-form relations
\begin{align}
\mathcal{T}_{c}^{a} \Omega_{b}^{c} = \tilde{\Omega}_{c}^{a} \mathcal{T}_{b}^{c} \notag\\
\mathcal{\bar{T}}_{\bar{c}}^{\bar{a}} \bar{\Omega}_{\bar{b}}^{\bar{c}} = \tilde{\bar{\Omega}}_{\bar{c}}^{\bar{a}} \mathcal{\bar{T}}_{\bar{b}}^{\bar{c}} \notag
\end{align}
which lead to the curvature relations
\begin{align}
\mathcal{T}_{b}^{a} R_{cd\bar{e}}^{b} = - \tilde{R}_{kl\bar{m}}^{a} \mathcal{T}_{d}^{l} \mathcal{\bar{T}}_{\bar{e}}^{\bar{m}} \mathcal{T}_{c}^{k} \notag\\
\mathcal{\bar{T}}_{\bar{b}}^{\bar{a}} R_{\bar{c}d\bar{e}}^{\bar{b}} = - \tilde{\bar{R}}_{\bar{k}l\bar{m}}^{\bar{a}} \mathcal{T}_{d}^{l} \mathcal{\bar{T}}_{\bar{e}}^{\bar{m}} \mathcal{\bar{T}}_{\bar{c}}^{\bar{k}} \notag
\end{align}
We search for further integrability conditions of these curvature
relations by repeated use of (\ref{equation16}) and
(\ref{equation17}). It is easy to obtain that covariant derivatives
of curvatures vanish so as to give that holomorphic pseudoduality is
between dual symmetric spaces with opposite curvatures. In fact,
this is due to the form of pseudoduality equations and always valid
no matter what the geometries of worldsheets and target spaces are

\subsection{Non-holomorphic Pseudoduality}\label{sec:MP}

It is allowed to mix the holomorphic vectors with anti-holomorphic ones on the pseudodual manifold. Therefore pseudoduality equations will be
\begin{align}
\tilde{X}_{\pm}^{a} = \pm \mathfrak{T}_{\bar{b}}^{a} \bar{X}_{\pm}^{\bar{b}} \label{equation18}\\
\tilde{\bar{X}}_{\pm}^{\bar{a}} = \pm \mathfrak{\bar{T}}_{b}^{\bar{a}} X_{\pm}^{b} \label{equation19}
\end{align}
where $\mathfrak{T} : T\mathcal{M}^{-} \longrightarrow
T\mathcal{\tilde{M}}^{+}$ and $\mathfrak{\bar{T}} : T\mathcal{M}^{+}
\longrightarrow T\mathcal{\tilde{M}}^{-}$ are mappings between mixed
tangent bundles switching the holomorhism. Taking exterior
derivatives of these equations and using (\ref{equation11}), and
corresponding Cartan's structure equations subsequently lead to
similar results as above $H = \tilde{H} = \tilde{\Gamma} = 0$ with
the constraint relations
\begin{align}
d\mathfrak{T}_{\bar{b}}^{a} - 2 \mathfrak{T}_{\bar{c}}^{a} \bar{\Lambda}_{\bar{b}}^{\bar{c}} + \tilde{\Lambda}_{c}^{a} \mathfrak{T}_{\bar{b}}^{c} = 0 \label{equation20}\\
d\mathfrak{\bar{T}}_{b}^{\bar{a}} - 2 \mathfrak{\bar{T}}_{c}^{\bar{a}} \Lambda_{b}^{c} + \tilde{\bar{\Lambda}}_{\bar{c}}^{\bar{a}} \mathfrak{\bar{T}}_{b}^{\bar{c}} = 0 \label{equation21}
\end{align}
where we defined $\bar{\Lambda}_{\bar{b}}^{\bar{a}} = \Gamma_{\bar{b}\bar{c}}^{\bar{a}} \bar{\Lambda}^{\bar{c}}$ and $\Lambda_{b}^{a} = \Gamma_{bc}^{a} \Lambda^{c}$. Taking exterior derivatives of (\ref{equation20}) and (\ref{equation21}) with the accompanying Cartan's second structural equations (\ref{equation12}) and (\ref{equation13}) yield the curvature relations
\begin{align}
\mathfrak{T}_{\bar{b}}^{a} R_{\bar{c}\bar{d}e}^{\bar{b}} = -\tilde{R}_{kl\bar{m}}^{a} \mathfrak{T}_{\bar{c}}^{k} \mathfrak{T}_{\bar{d}}^{l} \mathfrak{\bar{T}}_{e}^{\bar{m}} \notag\\
\mathfrak{\bar{T}}_{b}^{\bar{a}} R_{cd\bar{e}}^{b} = - \tilde{\bar{R}}_{\bar{k}\bar{l}m}^{\bar{a}} \mathfrak{\bar{T}}_{c}^{\bar{k}} \mathfrak{\bar{T}}_{d}^{\bar{l}} \mathfrak{T}_{\bar{e}}^{m} \notag
\end{align}
It is obvious that covariant derivatives of curvatures disappear by
means of (\ref{equation20}) and (\ref{equation21}), which bring
about the conclusion that manifolds at which sigma models based must
be dual symmetric spaces obeying the results found in
\cite{alvarez1, msarisaman1}. Consequently we determined that both
cases yield the same results that pseudoduality does not allow the
presence of torsions and manifolds are required to be symmetric
spaces with opposite curvatures. Pseudoduality could be performed
between (anti)holomorphic spaces or switch the holomorphicity.

\section{Isometries and Pseudoduality}\label{sec:IP}

The isometry group $G$ indicates that each point $X$ ($\bar{X}$) on the holomorphic (anti-holomorphic) part of the target space is moved so that metric of the target space remains unchanged \cite{ketov1, hull, bagger}. The infinitesimal action of $G$ is represented by \footnote{The finite form of the transformations are given by $X^{'a} = e^{L_{\lambda k}} X^{a}$ and $\bar{X}^{'\bar{a}} = e^{L_{\lambda k}} \bar{X}^{\bar{a}}$.}
\begin{align}
\delta X^{a} = \lambda^{A} k_{A}^{a} \label{equation22}\\
\delta \bar{X}^{\bar{a}} = \lambda^{A} \bar{k}_{A}^{\bar{a}} \label{equation23}
\end{align}
where $k_{A}^{a}$ ($\bar{k}_{A}^{\bar{a}}$) are the Killing vectors on the holomorphic (anti-holomorphic) part of $T \mathcal{M}^{\mathbb{C}}$, $A = 1, 2, ..., $dimG, and $\lambda^{A}$ are constant parameters. We adhere to the holomorphic isometry convention used in \cite{ketov1, lindstrom1}, $k = k(X)$ and $\bar{k} = \bar{k}(\bar{X})$. By means of $\mathcal{L}_{k} \partial\bar{\partial}K = 0$, It can readily be shown that Killing vectors required to have holomorphic isometry lead to Killing's equation
\begin{equation}
\nabla_{a} k_{A\bar{a}} + \nabla_{\bar{a}} \bar{k}_{Aa} = 0 \notag
\end{equation}
The holomorphic and anti-holomorphic components of the Killing vectors generate the distinct isometry algebras
\begin{equation}
k_{[A\bar{a}} k_{B]\bar{b}, \bar{a}} = f_{AB}^{C} k_{C\bar{b}} \ \ \ \ \ \ \ \ \bar{k}_{[Aa} \bar{k}_{B]b, a} = f_{AB}^{C} \bar{k}_{Cb} \notag
\end{equation}

We would like to see which conditions pseudoduality transformations impose on the holomorphic isometries. Thus we may write the following infinitesimal forms by means of (\ref{equation22}) and (\ref{equation23})
\begin{align}
\delta X_{\pm}^{a} = \lambda^{A} k_{A\pm}^{a} \label{equation24}\\
\delta \bar{X}_{\pm}^{\bar{a}} = \lambda^{A} \bar{k}_{A\pm}^{\bar{a}} \label{equation25}
\end{align}

Let us first analyze the holomorphic pseudoduality, and we work on the equations (\ref{equation14}) and (\ref{equation15}). Taking the infinitesimal forms yields that
\begin{align}
\tilde{\lambda}^{A} \tilde{k}_{A\pm}^{a} = \pm (\partial_{c} \mathcal{T}_{b}^{a}) \lambda^{A} k_{A}^{c} X_{\pm}^{b} \pm \mathcal{T}_{b}^{a} \lambda^{A} k_{A\pm}^{b} \notag\\
\tilde{\lambda}^{A} \tilde{\bar{k}}_{A\pm}^{\bar{a}} = \pm (\partial_{\bar{c}} \mathcal{\bar{T}}_{\bar{b}}^{\bar{a}}) \lambda^{A} \bar{k}_{A}^{\bar{c}} \bar{X}_{\pm}^{\bar{b}} \pm \mathcal{\bar{T}}_{\bar{b}}^{\bar{a}} \lambda^{A} \bar{k}_{A\pm}^{\bar{b}} \notag
\end{align}
where we defined $\partial_{c} \mathcal{T}_{b}^{a} \equiv \frac{\partial}{\partial X^{c}} \mathcal{T}_{b}^{a}$. We set $k_{A\pm} = k_{A\bar{b}} \bar{X}_{\pm}^{\bar{b}}$ and $\bar{k}_{A\pm} = \bar{k}_{Ab} X_{\pm}^{b}$, and corresponding pseudodual expressions to obtain
\begin{align}
\tilde{\lambda}^{A} \tilde{k}_{A\bar{b}}^{a} \mathcal{\bar{T}}_{\bar{c}}^{\bar{b}} \bar{X}_{\pm}^{\bar{c}} = \lambda^{A} (\partial_{b} \mathcal{T}_{c}^{a} k_{A}^{b} X_{\pm}^{c} + \mathcal{T}_{b}^{a} k_{A\bar{c}}^{b} \bar{X}_{\pm}^{\bar{c}}) \label{equation26}\\
\tilde{\lambda}^{A} \tilde{\bar{k}}_{Ab}^{\bar{a}} \mathcal{T}_{c}^{b} X_{\pm}^{c} = \lambda^{A} (\partial_{\bar{b}} \mathcal{\bar{T}}_{\bar{c}}^{\bar{a}} \bar{k}_{A}^{\bar{b}} \bar{X}_{\pm}^{\bar{c}} + \mathcal{\bar{T}}_{\bar{b}}^{\bar{a}} \bar{k}_{Ac}^{\bar{b}} X_{\pm}^{c}) \label{equation27}
\end{align}
Since $X_{\pm}$ and $\bar{X}_{\pm}$ are independent terms we find out that $\mathcal{T}_{c}^{a}$ and $\mathcal{\bar{T}}_{\bar{c}}^{\bar{a}}$ are constants and we choose them to be identity. Therefore we are left with the following relations
\begin{equation}
\tilde{\lambda}^{A} \tilde{k}_{A\bar{b}}^{a} = \lambda^{A} k_{A\bar{b}}^{a} \ \ \ \ \ \ \ \ \tilde{\lambda}^{A} \tilde{\bar{k}}_{Ab}^{\bar{a}} = \lambda^{A} \bar{k}_{Ab}^{\bar{a}} \label{equation28}
\end{equation}
with the pseudoduality equations reduced to
\begin{equation}
\tilde{X}_{\pm}^{a} = \pm X_{\pm}^{a} \ \ \ \ \ \ \ \ \tilde{\bar{X}}_{\pm}^{\bar{a}} = \pm \bar{X}_{\pm}^{\bar{a}} \label{equation29}
\end{equation}
It is intriguing that these are equivalent to well-known T-duality
transformations which have been obtained as a special case of
pseudoduality transformations in the presence of isometries. From
(\ref{equation28}) it is observed that isometries are related to
each other by
\begin{align}
\tilde{\lambda}^{A} \tilde{k}_{A}^{a} = \lambda^{A} (k_{A}^{a} + i\eta_{A}^{a}) \ \ \ \ \ \ \ \ \tilde{\lambda}^{A} \tilde{\bar{k}}_{A}^{\bar{a}} = \lambda^{A} (\bar{k}_{A}^{\bar{a}} - i\eta_{A}^{\bar{a}}) \label{equation30}
\end{align}
where $\eta$ is a real and constant vector on target space $\mathcal{M}^{\mathbb{C}}$.

In case of non-holomorphic pseudoduality, using expressions (\ref{equation24}) and (\ref{equation25}) we obtain
\begin{align}
\tilde{\lambda}^{A} \tilde{k}_{A\pm}^{a} = \pm (\partial_{c} \mathfrak{T}_{\bar{b}}^{a} \lambda^{A} k_{A}^{c} + \partial_{\bar{c}} \mathfrak{T}_{\bar{b}}^{a} \lambda^{A} \bar{k}_{A}^{\bar{c}}) \bar{X}_{\pm}^{\bar{b}} \pm \mathfrak{T}_{\bar{b}}^{a} \lambda^{A} \bar{k}_{A\pm}^{\bar{b}}\notag\\
\tilde{\lambda}^{A} \tilde{\bar{k}}_{A\pm}^{\bar{a}} = \pm (\partial_{c} \mathfrak{\bar{T}}_{b}^{\bar{a}} \lambda^{A} k_{A}^{c} + \partial_{\bar{c}} \mathfrak{\bar{T}}_{b}^{\bar{a}} \lambda^{A} \bar{k}_{A}^{\bar{c}}) X_{\pm}^{b} \pm \mathfrak{\bar{T}}_{b}^{\bar{a}} \lambda^{A} k_{A\pm}^{b}\notag
\end{align}
replacing $k_{A\pm}$ and $\bar{k}_{A\pm}$ lead to
\begin{align}
\tilde{\lambda}^{A} \tilde{k}_{A\bar{b}}^{a} \mathfrak{\bar{T}}_{c}^{\bar{b}} X_{\pm}^{c} = (\partial_{c} \mathfrak{T}_{\bar{b}}^{a} \lambda^{A} k_{A}^{c} + \partial_{\bar{c}} \mathfrak{T}_{\bar{b}}^{a} \lambda^{A} \bar{k}_{A}^{\bar{c}}) \bar{X}_{\pm}^{\bar{b}} + \mathfrak{T}_{\bar{b}}^{a} \lambda^{A} \bar{k}_{Ac}^{\bar{b}} X_{\pm}^{c}\label{equation31}\\
\tilde{\lambda}^{A} \tilde{\bar{k}}_{Ab}^{\bar{a}} \mathfrak{T}_{\bar{c}}^{b} \bar{X}_{\pm}^{\bar{c}} = (\partial_{c} \mathfrak{\bar{T}}_{b}^{\bar{a}} \lambda^{A} k_{A}^{c} + \partial_{\bar{c}} \mathfrak{\bar{T}}_{b}^{\bar{a}} \lambda^{A} \bar{k}_{A}^{\bar{c}}) X_{\pm}^{b} + \mathfrak{\bar{T}}_{b}^{\bar{a}} \lambda^{A} k_{A\bar{c}}^{b} \bar{X}_{\pm}^{\bar{c}}\label{equation32}
\end{align}
It is manifest that $\mathfrak{\bar{T}}_{b}^{\bar{a}}$ and $\mathfrak{T}_{\bar{b}}^{a}$ are constants and we show them by a subscript $\mathfrak{T}_{0}$. Thus the following relations remain
\begin{equation}
\tilde{\lambda}^{A} \tilde{k}_{A\bar{b}}^{a} (\mathfrak{\bar{T}}_{0})_{c}^{\bar{b}} = (\mathfrak{T}_{0})_{\bar{b}}^{a} \lambda^{A} \bar{k}_{Ac}^{\bar{b}} \ \ \ \ \ \ \ \ \tilde{\lambda}^{A} \tilde{\bar{k}}_{Ab}^{\bar{a}} (\mathfrak{T}_{0})_{\bar{c}}^{b} = (\mathfrak{\bar{T}}_{0})_{b}^{\bar{a}} \lambda^{A} k_{A\bar{c}}^{b} \label{equation33}
\end{equation}
with the corresponding pseudoduality equations
\begin{equation}
\tilde{X}_{\pm}^{a} = \pm (\mathfrak{T}_{0})_{\bar{b}}^{a} \bar{X}_{\pm}^{\bar{b}} \ \ \ \ \ \ \ \
\tilde{\bar{X}}_{\pm}^{\bar{a}} = \pm (\mathfrak{\bar{T}}_{0})_{b}^{\bar{a}} X_{\pm}^{b} \label{equation34}
\end{equation}
To find out the relations between isometries we make use (\ref{equation33}), which leads to
\begin{equation}
\tilde{\lambda}^{A} (\tilde{k}_{A}^{a} \mathfrak{\bar{T}}_{0}) =
\lambda^{A} (\mathfrak{T}_{0})_{\bar{b}}^{a}  (\bar{k}_{A}^{\bar{b}}
- i \eta_{A}^{\bar{b}}) \ \ \ \ \ \ \ \ \tilde{\lambda}^{A}
(\tilde{\bar{k}}_{A}^{\bar{a}} \mathfrak{T}_{0}) = \lambda^{A}
(\mathfrak{\bar{T}}_{0})_{b}^{\bar{a}} (k_{A}^{b} + i \eta_{A}^{b})
\label{equation35}
\end{equation}
where $\eta$ is a real and constant vector.

We see that holomorphic isometries reduce pseudoduality equations to
(\ref{equation29}) in case of holomorphic pseudoduality (or
(\ref{equation34}) in case of anti-holomorphic pseudoduality), and
therefore isometries are preserved. The resultant expressions are
the familiar T-duality transformations, which is a special case of
pseudoduality in case of isometries. Therefore, it comes out that if
certain conditions are imposed on pseudoduality transformations,
they can be reduced to other well-known duality equations in
literature. An interesting case could be the reduction of
pseudoduality to the mirror symmetry transformations which requires
a thorough study of topological sigma models, which will not be
discussed here.

\section{Discussion}\label{sec:D}

We have seen that extension of $N = (1, 1)$ sigma models to $N = (2, 2)$ case with the constraints given by (\ref{equation1}) results in the target space a K\"{a}hler manifold, which is the complexified target space. All the structures on target space can be split into holomorphic and anti-holomorphic parts using projection operator (\ref{equation7}) once complexification comes true. It has been pointed out that pseudoduality stands out in a well represented holomorhic (\ref{equation14}) and (\ref{equation15}) or anti-holomorphic (\ref{equation18}) and (\ref{equation19}) forms. They all give the constraints that torsions related to each part has to vanish. Also on the pseudodual space riemannian connection disappears, which proves that all the points on the target space are mapped into a single point where riemann normal coordinates hold.

It is shown that holomorphic isometries on both $\mathcal{M}$ and
$\tilde{\mathcal{M}}$ are preserved, and pseudoduality equations are
further restricted to have a constant mapping from one space to
another one, which gives rise to T-duality. It is obvious that if
transformation contains superfields then isometries has to contain
both holomorphic and anti-holomorphic superfields, which is the
origin of obtaining T-duality transformations from the reduction of
pseudoduality. Therefore, it turns out that pseudoduality
transformations incorporate other symmetries and can be reduced once
relevant conditions are imposed.

\section*{Acknowledgments}

I would like to thank O. Alvarez for his comments, helpful
discussions, and reading an earlier draft of the manuscript.

\bibliographystyle{amsplain}

\begin{thebibliography}{9}

\bibitem{ketov1}
  S. V. Ketov,
  \emph{Quantum Non-Linear Sigma-Models: From Quantum Field Theory to
Supersymmetry, Conformal Field Theory, Black Holes and Strings,
Texts and Monographs in Physics}, Springer, Berlin, Germany, 2000.

\bibitem{alvarez1}
  O. Alvarez,
  \emph{Pseudoduality in Sigma Models},
  Nucl.Phys. B638 (2002) 328- 350, \href{http://www.arxiv.org/pdf/hep-th/0204011/}{hep-th/0204011/}.

\bibitem{alvarez2}
  O. Alvarez,
  \emph{Target space pseudoduality between dual symmetric spaces},
  Nucl. Phys. B582 (2000) 139, \href{http://www.arxiv.org/pdf/hep-th/0004120/}{hep-th/0004120/}.

\bibitem{curtright1}
  T. Curtright and C.Zachos,
  \emph{Currents, charges, and canonical structure of pseudochiral models},
  Phys. Rev. D49 (1994) 5408-5421, \href{http://www.arxiv.org/pdf/hep-th/9401006/}{hep-th/9401006/}.

\bibitem{ivanov}
 E. A. Ivanov,
 \emph{Duality in d = 2 sigma models of chiral field with anomaly}, Theor. Math. Phys. 71 (1987) 474-484.

\bibitem{msarisaman1}
 M.Sarisaman,
\emph{Pseudoduality In Supersymmetric Sigma Models}, Int.J.Mod.Phys.A25:2997-3023,2010,
\href{http://arxiv.org/abs/0904.4408/}{hep-th/0904.4408/}.

\bibitem{nakahara}
 M. Nakahara,
 \emph{Geometry, topology and physics}, Institute of Physics Publishing, 1990.

\bibitem{zumino}
B.Zumino,
 \emph{Supersymmetry And Kahler Manifolds}, Phys. Lett. B 87, 203 (1979).

\bibitem{lindstrom1}
 U. Lindstr\"{o}m,
\emph{Supersymmetry, a Biased Review},
\href{http://arxiv.org/abs/020401v2/}{hep-th/020401v2/}.

\bibitem{hull}
C. M. Hull, A. Karlhede, U. Lindstrom and M. Rocek,
\emph{Nonlinear Sigma Models And Their Gauging In And Out Of Superspace}, Nucl. Phys. B 266, 1 (1986).

\bibitem{bagger}
J. Bagger,
\emph{Coupling the Gauge Invariant Supersymmetric Nonlinear Sigma Model to Supergravity}, Nucl. Phys. B211, 302 (1983).


\end{thebibliography}

\end{document}